\documentclass[preprints,article,accept,pdftex,moreauthors]{Definitions/mdpi} 
\firstpage{1} 
\pubvolume{1}
\issuenum{1}
\articlenumber{0}
\pubyear{2022}
\copyrightyear{2022}
\datereceived{} 
\dateaccepted{} 
\datepublished{} 
\hreflink{https://doi.org/} % If needed use \linebreak

\Title{Propagation of cosmic rays in plasmoids of AGN jets - implications for multimessenger predictions}

\TitleCitation{Propagation of cosmic rays in AGN jets}

\Author{Julia Becker Tjus $^{1,2}$\orcidA{}, Mario H\"orbe$^{1,2}$\orcidB{}, Ilja Jaroschewski$^{1,2}$\orcidC{}, Patrick Reichherzer$^{1,2,3}$\orcidD{}, Wolfgang Rhode $^{4}$\orcidE{}, Marcel Schroller $^{1,2}$\orcidF{}, Fabian Schüssler$^{3}$\orcidG{}}

\AuthorNames{Becker Tjus, H\"orbe, Jaroschewski, Reichherzer, Rhode, Schroller, Sch\"ussler}

\AuthorCitation{Becker Tjus, J.; H\"orbe, M.; Jaroschewski, I.; Reichherzer, P.; Rhode, W.; Schroller, M. and Sch\"ussler, F.}
\address{%
$^{1}$\quad Ruhr-Universit\"at Bochum, Theoretische Physik IV: Plasma-Astroteilchenphysik, Universit\"atsstrasse 150, 44801 Bochum, Germany\\
$^{2}$ \quad Ruhr Astroparticle and Plasma Physics Center (RAPP Center), Ruhr-Universit\"at Bochum, 44780 Bochum, Germany\\
$^{3}$ \quad IRFU, CEA, Université Paris-Saclay, F-91191 Gif-sur-Yvette, France\\
$^{4}$ \quad Department of Physics, TU Dortmund University, 44221 Dortmund, Germany}

\corres{Correspondence: julia.tjus@rub.de}

\abstract{After the successful detection of cosmic high-energy neutrinos, the field of multiwavelength photon studies of active galactic nuclei (AGN) is entering an exciting new phase. The first hint of a possible neutrino signal from the blazar TXS\,0506+056  leads to the anticipation that AGN could soon be identified as point sources of high-energy neutrino radiation, representing another messenger signature besides the well-established photon signature. To understand the complex flaring behavior at multiwavelengths, a genuine theoretical understanding needs to be developed. These observations of the electromagnetic spectrum and neutrinos can only be interpreted fully when the charged, relativistic particles responsible for the different emissions are modeled properly. The description of the propagation of cosmic rays in a magnetized plasma is a complex question that can only be answered when analysing the transport regimes of cosmic rays in a quantitative way.  In this paper, we therefore present a quantitative analysis of the propagation regimes of cosmic rays in the approach that is most commonly used to model non-thermal emission signatures from blazars, i.e.\ the existence of a high-energy cosmic-ray population in a relativistic plasmoid traveling along the jet axis. In this paper, we show that in the considered energy range of high-energy photon and neutrino emission, the transition between diffusive and ballistic propagation takes place, significantly influencing not only the spectral energy distribution but also the lightcurve of blazar flares.
}


\begin{document}
\section{Introduction}
Active galactic nuclei (AGN) are among the most enigmatic objects in the Universe.
With luminosities in excess of $10^{37}$\,W ($10^{44}$\,erg\,s$^{-1}$), they represent the most luminous, continuous sources of radiation. With their central supermassive black holes (SMBHs), they provide an environment that can help us to understand black holes at work. The question of how energy is transferred from the black hole and/or the accretion disk to launch  gigantic radio jets is still largely unsolved and subject to ongoing research.
AGN also constitute one of the few sites in the Universe that provide enough energy on total to serve as a candidate for the observed flux of ultra-high energy cosmic rays (UHECRs) and might be a key to understand particle acceleration up to macroscopic energies, see e.g.\ \cite{a7:beckertjus_multimessenger,biermann2009} for  summaries. As one of the very few source classes, active galaxies provide an astrophysical, extreme environment which might be suited to accelerate particles to an incredible amount of $10^{20}$\,eV \citep[see e.g.][]{a7:biermann_strittmatter1987}. These sources are therefore also considered to contribute to the diffuse astrophysical high-energy neutrino flux as measured by IceCube at Earth \citep{a7:icecube2013}. In particular, the sub-class of blazars is known for their strong short- and long-term variability, especially (but not exclusively) at gamma-ray energies. 
In the literature, blazar jets are often discussed to  be dominated by leptonic particle processes which fit observational quiescent data of blazar spectral energy densities \citep[SEDs; see][]{a7:reynolds1996matter_leptonJet, a7:wardle1998electron_leptonJet, a7:potter2012synchrotron}. Yet, the understanding of the complex, time-variable structure of the SEDs is still far from being understood in all detail. Also, the picture of an electron-positron plasma in the jet is not the only possibility, and an electron-proton (hadron) plasma is certainly a realistic option. Thus, in the past decades, AGN jets and particularly blazar emissions have also been argued to naturally contain hadronic components which would not only contribute to the emission of electromagnetic radiation in blazars \citep{a7:bottcher2013leptonic_inferHadronsInEMemission}, but also lead to the production of secondary high-energy neutrinos 
\cite{ownpub_a7:bbr2005, ownpub_a7:jbt2014,a7:murase2012blazars_blazarsGenerateUHECR}. The general detection of up to PeV high-energy neutrinos of astrophysical origin observed by IceCube  \citep{a7:icecube2013} has started to shed more light on the  non-thermal high-energy Universe. The existence of such a diffuse high-energy neutrino flux implies that there must be one or  more source classes that produce high-energy neutrinos via (photo-)\-hadronic interactions. From the distribution of events in the projected sky, it is clear that the flux is not focused in the Galactic plane, and that it is therefore likely to contain a significant extragalactic component, see e.g.\ \citet{a7:beckertjus_merten2020} for a summary. In the past few years, several possible associations of neutrinos with astrophysical objects have been identified. Each of these is at a $\sim 3\upsigma$ level so far and serves as first evidence. 
A first hint for an association of a high-energy neutrino with a blazar comes from the source TXS\,0506+056. In September 2017, a neutrino of $\sim300$\,TeV could be associated with an exceptional gamma-ray flare at GeV energies from this source. This gamma-neutrino correlation can be estimated to have a  significance of $3\upsigma$ by combining data of the IceCube neutrino observatory and Fermi-LAT \cite{ownpub_a7:txs}. This detection initiated a dedicated search for neutrino clustering in the $\sim 10$\,years of IceCube data around the position of TXS\,0506+056, with the result that there was an enhanced flux of neutrinos in a half-year period from September 2014 to March 2015, also with a statistical significance of $\sim 3\upsigma$ of being incompatible with the background hypothesis \citep{txs_science2018b}.

In order to explain the neutrino signatures detected from the direction of TXS\,0506+065, hadronic jet models have been applied \cite[e.g.][]{a7:halzen2019,a7:gao2019,a7:rodrigues2019,a7:debruijn2020}  and \cite{ownpub_a7:model2020}. The two potential flares appear quite different in their evolution: the neutrino signal above the atmospheric background detected in 2014/2015 lasted about 100\,days and consisted of about $8-18$ neutrinos with energies of $10-100$\,TeV, and the gamma-ray light curve at GeV is in a minimum state \citep{txs_science2018b}. The 2017 detection is based on one single high-energy neutrino with extreme energy ($\sim300$\,TeV). A gamma-ray flare was observed in coincidence with the arrival of the neutrino. It has been noted by \citet{Kun2021}, however, that at the exact time of the neutrino detection, even here the GeV gamma-ray flux was at a local minimum and only rose to a high emission state shortly after the neutrino detection. It was argued in \cite{Kun2021} that this observation is consistent with a model for which the neutrinos are produced in a high-density medium in which gamma-rays are absorbed, which either becomes less dense with time or for which the gamma-rays take some time to cascade down to GeV energies before escaping.   

By now, there are tens of possible associations of neutrinos and AGN jets \citep{a7:Kadler2016,a7:Franckowiak2020,a7:Giommi2020}. With IceCube in continued operation, this number will further increase in the upcoming years, with the expectation to finally confirm at least some of these sources at the $>5$~sigma level to be neutrino emitters. One common conundrum in all of the different high-energy neutrino detections (diffuse and potential sources)  is the lack of TeV, but even GeV gamma-ray emission. Hadronic interactions that are responsible for neutrino production inevitably lead to the co-production of high-energy gamma rays. The reason is that neutrinos are produced from the subsequent decay of charged pions and kaons, which in turn are produced in a fixed ratio of neutral pions and kaons, leading to the production of gamma rays with an energy threshold at the mass of the pion, $E_{\gamma,\,\min}=m_{\pi_0}\,c^2/2=70$\,MeV. Comparing the detected diffuse neutrino flux with the measured extragalactic, diffuse component of gamma rays leads to the conjecture of a source environment that must absorb the gamma rays at energies $>$~GeV \citep{a7:2016PhRvL.116g1101M,
a7:ahlers2015}. The absorption of photons can be due to a strong accretion disk, as it has been discussed in e.g.\ \citet{a7:brodatzki2011} for the case of TeV emission. The potential neutrino source fluxes from the 2014/2015 TXS signature also indicates that, in order to match the observed gamma-ray flux, it must be diminished significantly at $>$ GeV energies to make the neutrino production model work. Such a model of gamma-ray absorption can even be used as a tracer in the searches for associations of high-energy neutrinos with blazars. That is, rather than searching for an enhanced gamma-ray flux, the neutrinos can actually arrive at times of reduced gamma-ray activity \cite{Kun2021}. Such a scenario can be produced for regions of extreme gas or photon densities. In the first case, the photons will interact with the dense gas via Compton scattering. In the second case, gamma-gamma interactions will lead to electromagnetic cascades. It is clear that in both scenarios, the energy of the high-energy gamma rays must be visible in the end at other wavelengths. If these environments only become transparent at MeV energies as suggested by e.g.\ \citet{a7:halzen2019}, this is not observable now, as a dedicated mission for MeV detection of the gamma-ray sky is currently missing. Future missions like e-Astrogam, MeVCube, or AMEGO  will shed more light on these questions. At this point, the theoretical models are being challenged by GeV-TeV measurements, which indicate that there is no significant increase in the energy output connected to the neutrinos.

In general, the modeling of the steady-state emission, but even more the modeling of the flares is complex and requires a complete consideration of the jet physics, including different scenarios for the acceleration region, gas and photon targets, as well as the magnetic field structure. The latter is highly important for the proper modeling of the diffusive cosmic-ray transport, which is relevant for the evolution of the flares, both for the leptonic and hadronic signatures \cite{ownpub_a7:model2020}.
Quantitative theoretical modeling is necessary to establish a physical connection of the neutrinos to the blazars. Mere directional coincidence is not enough, because the angular uncertainty of the neutrino events is larger than $1^{\circ}$, making source identification difficult without theoretical input.

Models of high-energy neutrino and electromagnetic up to gamma-ray emission in the jets of AGN cover a variety of scenarios and parameter spaces. Blazars are known to be highly variable across the electromagnetic spectrum, with a variety of models put forward to explain these flares. Such models include, among others, particle acceleration via internal shocks in the jet 
\citep[see e.g.][and \citealp{a7:eichmann2012}]{a7:biermann_strittmatter1987, ownpub_a7:bbr2005, ownpub_a7:jbt2014} and reconnection driven plasmoids \citep[see e.g.][and \citealp{ownpub_a7:model2020}]{a7:giannios2013reconnection_PlasmoidGamma, a7:morris2019feasibility_paulsPlasmoid}. These latter relativistic and compact structures have been discussed in the literature since the 1960s \citep{a7:rees1966,a7:vanderlaan1966}. What we refer to as \textit{plasmoid} is often called \textit{blob} in the literature, meaning compact, dense structures traveling with relativistic speeds along the jet axis. The term \textit{plasmoid} is preferred here, as it is typically used in the context of the plasmoid creation via magnetic reconnection events. In this scenario, the injection of a relativistic plasma into the system (here the AGN jet) can lead to reconnection events that, under certain circumstances, lead to the plasmoid instability that breaks down the streaming plasma into small \textit{blobs}, i.e.\ the plasmoids. In this scenario, charged particles can be pre-accelerated in the reconnection events. While non-relativistic reconnection is limited to below-knee energies \citep{a7:lyutikov-komissarov-sironi-porth-2018}, relativistic reconnection can be much more efficient \citep{a7:sironi-spitkovsky-2014}, also by further acceleration via a Fermi second-order process when the particles scatter in between the plasmoids.  Such an acceleration scenario  can  solve the long-standing injection-problem. They also justify the assumption that the cosmic-ray population is distributed homogeneously in the plasmoid, as the turbulent field in the plasmoid is used to isotropize the direction of the incoming particles. The assumption of a homogeneously distributed population is implicit in those models that do not resolve the blob, but work with timescales. In test-particle simulations, it is a reasonable approach to start with a homogeneous distribution as we will do in this paper.

The modeled hadronic component of  proton-proton interaction was discussed in \citet{a7:eichmann2012}. 
The modeling of leptonic and (lepto-)hadronic emission in \citet{a7:christie2018radiative_EquilibriumPlasmoids_B1G, a7:keivani2018multimessenger_blazarNuGamma} for the case of TXS\,0506+056. Other flare models based on external factors include gas clouds entering the base of jets \citep{a7:dar1997hadronic_cloudEntersJet, a7:araudo2010gamma_cloudEntersJet, a7:zacharias2019extended_cloudAblationByJet} and jets of former binary AGN drilling through their own dust tori and/or accretion disks after being redirected by the merger of their host black holes \citep{a7:gergely2009spin_flip}. Such scenarios of high density all tend to be more hadron-dominated due to the nature of their occurrence. 

\begin{figure}
  \centering
  \includegraphics[width=0.7\textwidth]{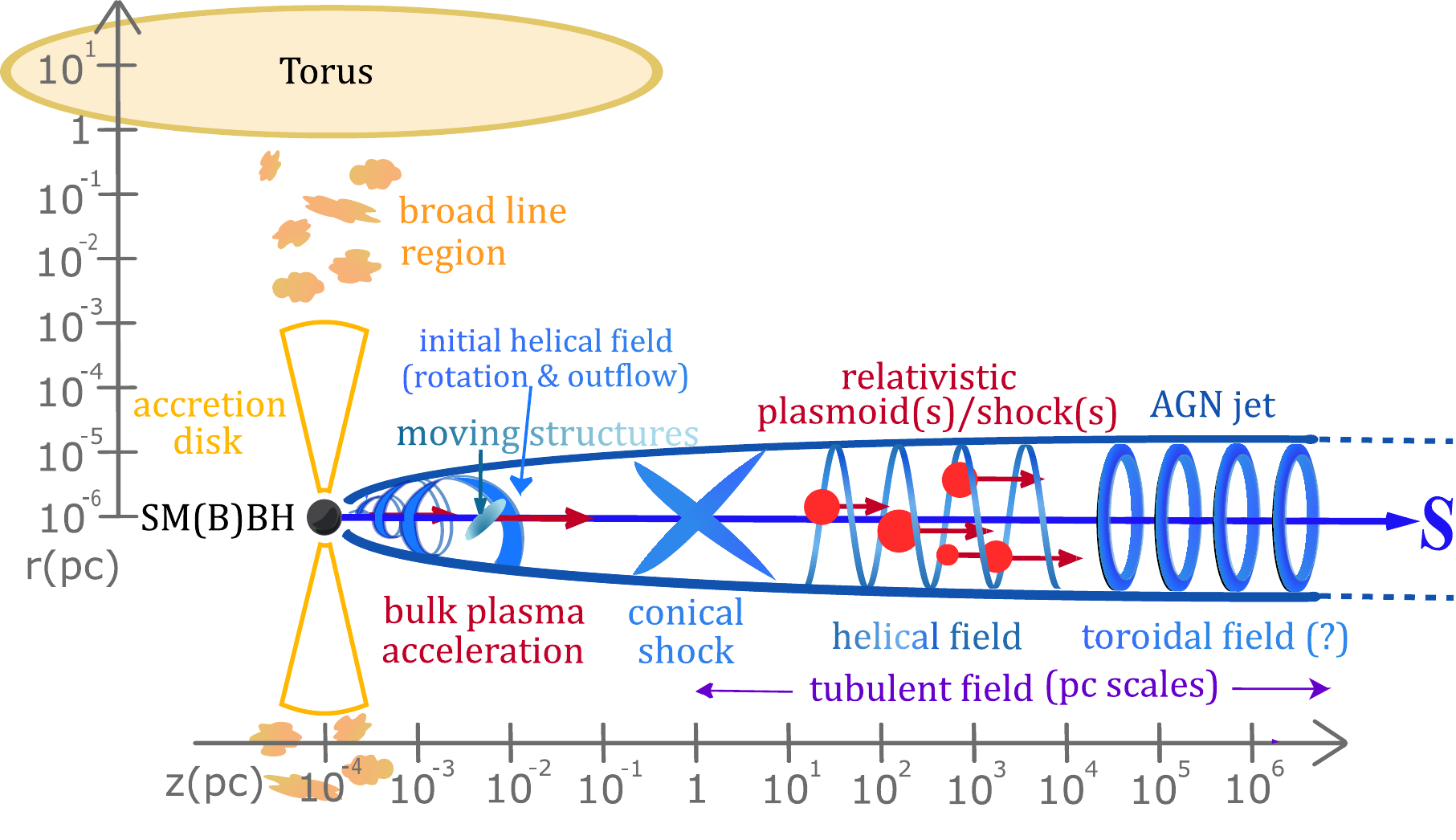}
  \caption{Structure of an AGN with a jet. Yellow/red components are targets for cosmic-ray interactions with gamma rays or gas. Blue colors indicate the likely structure of the magnetic field along the jet. Particle acceleration happens either at the shock fronts or in context with the relativistic plasmoids.}
 \label{fig_a7:agnjet}
\end{figure}

Figure~\ref{fig_a7:agnjet} shows a sketch of an AGN with a jet with the photon and/or gas targets. The structure of the large-scale magnetic field is shown in blue. A turbulent component of the magnetic field exists as well and is not drawn in the figure, but only indicated by purple text. The structure of the gas/photon fields is highly relevant for the particle interactions and therefore needs to be included in the models in three dimensions. The same is true for the magnetic field structure as the synchrotron radiation is sensitive to the direction of the field. For a diffusive transport description, it is also highly relevant as it is often assumed that the propagation is dominantly along the field lines with a smaller component perpendicular to the field. In fact, the perpendicular diffusion coefficient can even dominate the picture if the turbulence level is $\delta B/B > 1$, something that is certainly possible in these extreme environments.

Current state-of-the-art numerical codes include many of the necessary features. The codes and their most important properties are summarized in Table~\ref{a7_table:agn_codes}. The models are typically designed to numerically solve the transport equation including loss processes from which the secondary particle radiation from electrons and protons can be calculated. A loss term for the particle transport is usually included via a timescale, i.e.\ a term $-n/\tau$, where $n$ is the particle density and $\tau$ is the characteristic timescale. In \cite{a7:cerruti2015,a7:dimitrakoudis2012,boettcher2013a}, the timescale is chosen to be the ballistic one, $\tau=c/R$, thus being energy-independent.
In \citet{a7:gao2017}, it is argued that the propagation is of diffusive nature so that the escape time of particles is assumed to be a factor of $10$ times longer than the ballistic one, but still assumed to be energy-independent.
All of these models are designed to model propagation and particle interaction in blazars. Due to the strong variability of the sources, it can be deduced that the signal must come from the very compact region on the order of $10^{14}$~m. This makes the plasmoid model favorable over an approach of shock acceleration and explains why all codes focus on such an approach. 

A new code for propagation of particles in relativistic plasmoids has been developed recently \cite{ownpub_a7:model2020}. This new framework has been derived from the public transport code CRPropa 3.1 \cite{merten2017}. The advantage with this numerical approach is that ballistic propagation can be performed in a test particle approach, thus not relying on the simplifying assumption of time-scales. Further, a second transport framework is integrated in CRPropa 3.1, which is the solution of the transport equation via the approach of Stochastic Differential Equations (SDEs). This approach enables to solve the transport equation via pseudo-particle propagation, which makes it compatible to be used in the ballistic test-particle propagation of CRPropa. The SDE approach is designed to include a full diffusion tensor, which is also an improvement when compared to other codes. A CRPropa modification presented in \citet{ownpub_a7:model2020}  makes use of the modular structure of CRPropa to create a propagation environment of a plasmoid traveling along the jet axis. The photon field of a thin accretion disk is implemented at the foot of the jet for gamma-gamma and proton-gamma interactions. Technically, the propagation is done in the reference frame of the plasmoid and then transferred into the observer's frame. The plasmoid itself contains a plasma with a constant density $n_{\rm plasma}$ which is considered as a target for cosmic-ray interactions as well. The magnetic field in which the particles propagate was assumed to be of purely turbulent nature of Kolmogorov type in \citet{ownpub_a7:model2020}. Due to the modular structure of the code, it can easily be changed to include a regular component as well, to change the nature of the turbulence, etc.

In this paper, we are putting the spotlight on the propagation regime in the plasmoids of blazars. In Section \ref{space:sec}, we quantify the energy at which a transition between diffusive and ballistic propagation is happening and what consequences such a transition has for the description of SEDs and lightcurves of blazars.  In Section \ref{time:sec}, we investigate the influence of a first phase of ballistic propagation before the limit of diffusion is reached in time and discuss the necessity to go from a diffusive approach to the description via the telegraph equation. In Section \ref{sec:simulation}, we perform first test simulations to investigate a possible difference in the flaring behavior in the diffusive vs.\ ballistic description.
Conclusions and outlook are given in Section \ref{conclusions:sec}.

\begin{table}
\begin{small}
\begin{tabular}{
|>{\RaggedRight}p{0.15\textwidth-2\tabcolsep}
|>{\Centering}p{0.12\textwidth-2\tabcolsep}
|>{\Centering}p{0.14\textwidth-2\tabcolsep}
|>{\Centering}p{0.17\textwidth-2\tabcolsep}
|>{\Centering}p{0.14\textwidth-2\tabcolsep}
|>{\Centering}p{0.13\textwidth-2\tabcolsep}
|} \hline %\hhline{|~|-|-|-|-|-||-|}
     Code & AM$^3$ & PARIS & ATHE$\nu$A & B\"ottcher & CRPropa\\ \hline
   Reference & \citet{a7:gao2017} &\citet{a7:cerruti2015} & \citet{a7:dimitrakoudis2012}   &\citet{boettcher2013a} & \citet{ownpub_a7:model2020} \\ \hline
   Transport equation  & yes & yes & yes & yes & yes \\ \hline 
   Ballistic  & no & no & no & no & yes\\ \hline
   steady state  & yes &yes & yes & yes& yes\\ \hline 
  time dependent  & yes & no & yes & no& yes \\ \hline
  B-field  & turbulent (isotropic) & turbulent (isotropic) & turbulent (isotropic) & turbulent (isotropic) & turbulent (isotropic), regular (helical)\\ \hline 
  Diffusion  & 1-dim & 1-dim & 1-dim & 1-dim &3-dim\\ \hline
 Photohadron &yes&yes&yes&yes&yes\\\hline
 Hadron-hadron &no&no&no&no&yes\\\hline
 \end{tabular}
 \caption{Basic properties of state-of-the-art blazar propagation codes.\label{a7_table:agn_codes}}
\end{small}
\end{table}

%%%%%%%%%%%%%%%%%%%%%%%%%%%%%%%%%%%%%%
\section{The space-domain: diffusive vs.\ ballistic propagation \label{space:sec}}
%%%%%%%%%%%%%%%%%%%%%%%%%%%%%%%%%%%%%%
As discussed above, propagation of charged particles in a turbulent (plus regular) magnetic field can be of fundamentally different nature depending on the astrophysical setting, in particular concerning the parameters of the particle energy $E$, the ratio of the turbulent to regular magnetic field $\delta B/B$, and the correlation length of the field $l_c$ as the lower boundary for the deterministic description of the magnetic field lines. In \citet{reichherzer2020}, five propagation regimes are quantified with respect to the particle's reduced rigidity $\rho=r_g/l_c$, with $r_g=E/(c\,q\,B)$ as the relativistic gyro radius and $l_c\approx l_{\max}/5$ as the correlation length. Here, $l_{\max}=2\pi/k_{\min}$ is the maximum scale of the magnetic turbulence spectrum, connected to the lowest wave number $k_{\min}$, defined by the turbulence injection scale. Diffusive propagation corresponds to the 
\textit{resonant-scattering regime} (RSR). This regime is only valid for particles that can scatter with the entire spectrum of wavelengths $k_{\min}<k<k_{\max}$, where $k_{\max}=2\pi/l_{\min}$ is the dissipation scale. The general scheme of resonant scattering then breaks down toward the lowest and highest reduced rigidities. At the lower boundary, when scattering does not happen with the entire angular spectrum anymore, mirroring occurs more and more often, altering the diffusion coefficient. It is expected that this regime is not relevant for particles in AGN plasmoids, as the gyro radius of $\sim$~TeV-PeV particles that are considered here in magnetic fields of $\sim$~G strength fulfills the boundary condition $\rho>l_{\min}/(\pi\,l_c\,\delta B/B)$ \cite{reichherzer2020}. The situation is different toward large reduced rigidities, for  which particles propagate in a \textit{quasi-ballistic} way: for gyro radii that start to reach the correlation length of the system, meaning for plasmoids also coming closer to the actual size of the system, only a few gyrations are performed by the particles before leaving the source. This is happening close to the Hillas limit of the source. It is clear that the number of gyrations then does not suffice for a diffusive description. 

The quasi-ballistic regime becomes relevant at a reduced rigidity of $\rho = r_g/l_c \gtrsim  5/(2\pi)$ \cite{reichherzer2020}.
Inserting the relativistic gyro radius, the energy at which the ballistic regime becomes dominant is given as
\begin{equation}
    E \gtrsim \frac{5}{2\pi} \cdot l_c \cdot c \cdot q \cdot B\,.
    \label{eq:for_Hillas_plot}
\end{equation}
For a given parameter set of magnetic field strengths, coherence lengths of the turbulence, and particle energies, Eq.\ (\ref{eq:for_Hillas_plot}) can be applied to determine in what regimes the particles are propagating in typical astrophysical sources of cosmic rays \cite{Reichherzer2021ICRC}. Normalized to a standard set of parameters, the equation becomes
\begin{equation}
    E\gtrsim Z \cdot \left( \frac{l_c}{10^{11}\,\mbox{m}}\right) \cdot \left(\frac{B}{0.42\,\mbox{G}}\right)\cdot 10^{15}\,\mbox{eV}\,.
\end{equation}
This implies that for protons ($Z=1$) in a source region with a parameter set $l_c=10^{11}$~m and $B=0.42$~G, diffusive propagation is happening below energies of $10^{15}$~eV, ballistic propagation needs to be applied above $10^{15}$~eV. For other parameter combinations, this transition energy can be calculated accordingly, always with ballistic propagation above the energy, diffusive transport below.

\begin{figure}
    \centering
    \includegraphics[width=0.8\textwidth]{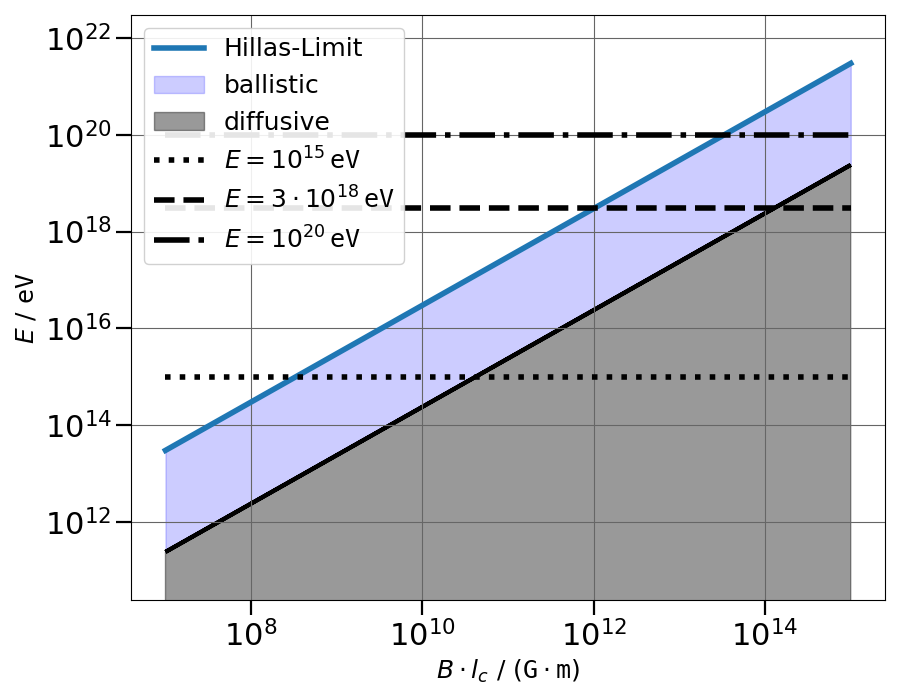}
    \caption{Illustration of the transition between the quasi-ballistic regime (blue) and the resonant scattering regime (grey) in dependence of the magnetic field strength, correlation length of the magnetic field and particle energy. In a simplified approach, it is assumed here that particles in the quasi-ballistic regime propagate ballistically and in the resonant scattering regime diffusively. 
    The position of the diagonal line represents the transition energy from diffusive (below) to ballistic (above) for protons ($Z=1$), determined by Eq.\ \ref{eq:for_Hillas_plot}.
    Diffusive propagation is expected for the result of a parameter combination below the line, while ballistic propagation lies above the line.}
    \label{fig:Plasmoid}
\end{figure}

Figure \ref{fig:Plasmoid} shows the energy limits for protons as a function of the product $B\cdot l_c$. 
The grey shaded area illustrates diffusive and the blue area ballistic propagation, with the transition between the resonant scattering regime and  the quasi-ballistic regime indicated as the solid line in between, following Eq.\ (\ref{eq:for_Hillas_plot}).   
The area of ballistic propagation in blue is bounded by the maximum possible proton energies according to the Hillas-Limit in each parameter space. The horizontal lines indicate the energies for the knee (dotted, $10^{15}$~eV), ankle (dashed, $10^{18.5}$~eV), and maximum observed energy (dashed-dotted, $10^{20}$~eV).

The parameter space covered by the plasmoids is approximated to be in the range $10^{10}$~m~$<l_c<10^{14}$~m. This range is based on the assumption that the plasmoids have a radius on the order of $R\sim 10^{12}-10^{16}$~m, using a correlation length of $l_c=0.01\cdot R$ as described above.
As the plasmoids are launched at the foot of the jet, magnetic fields are large, on the order of $10^{-3}$~G~$<B<10$~G. %\Julia{Referenz?}. 
What we want to understand is how the propagation of particles needs to be performed to describe the multimessenger emission from AGN in the plasmoid-model. The energy range of interest for high-energy photons reaches from GeV energies up to approximately $10^{16}$~eV, neutrino detection happens in an energy range corresponding to proton energies of approximately $2\cdot 10^{13}$~eV to $10^{17}$~eV. 
Figure \ref{fig:Plasmoid} shows the relevant parameter space, displayed as $B\cdot l_c$ on the x-axis, a fraction will be diffusive (grey area) at lower energies. The high-energy part before reaching the Hillas limit (colored, thick line) needs to be performed in the ballistic limit (blue area). A first  extreme example is a combination $B\cdot l_c=10^{8}$~G~m, where diffusive propagation happens up to $\sim 10^{12.5}$~eV, ballistic propagation up to the Hillas limit at around $10^{14.5}$~eV. These would be sources with a relatively low acceleration limit, as the combination of $R$ and $B$ only allows for maximum energies below the knee. More realistic parameter combinations that would allow the sources to reach the maximum observed energy would be a combination of $B\cdot l_c=10^{14}$~G~m. In this case, diffusive propagation needs to be performed up to $10^{18.5}$~eV, ballistic propagation needs to be performed up to the Hillas limit at $10^{20.5}$~eV.

This result has immediate consequences on the observed energy spectra: a break in the energy behavior of the timescales applied in simplified transport equation approaches, where the term $-n/\tau_{\rm esc}$ describes the escape, is expected to be observed: For ballistic transport, the timescale needs to be chosen as a constant value, $\tau_{\rm esc}^{\rm ballistic}=R/c$, while it becomes energy dependent in the case of diffusive propagation, $\tau_{\rm esc}^{\rm diffusive}=R^2/\kappa \propto \sqrt{R}\,E^{-\delta}$, with $\kappa=\kappa_0\cdot E^{\delta}$ as the diffusion coefficient, for which the energy dependence can be  approximated as a power-law behavior with an index $\delta$ that depends on the underlying magnetic turbulence, in this description being in the limit of quasi-linear theory, i.e.\ $\delta B/B \ll 1$. Such a change in the timescale directly induces a change in the shape of the spectral energy distribution: applying the leaky box model, the emitted proton spectrum follows approximately $n(E)\sim Q(E)\cdot \tau_{\rm esc} (E)$. Those secondary photons and neutrinos that are induced by (photo-)hadronic interactions basically mirror that behavior in the energy range above the threshold for the process, so that even these are in first approximation proportional to the escape time and the primary injection spectrum $Q(E)$, $n_{\gamma,\nu}\propto Q(E)\cdot \tau_{\rm esc}$. Assuming an injection spectrum $Q(E)\propto E^{-2.3}$ and Kolmogorov-type turbulence, $\tau_{\rm esc}^{\rm diffusive}\propto E^{-0.3}$, the SED is expected to behave as
\begin{equation}
    n_{\gamma,\nu}(E)\propto\left\{\begin{array}{ll} E_{\gamma, \nu}^{-2.6}& E<E_{\rm transition}^{\gamma,\nu}\\
    E_{\gamma, \nu}^{-2.3}&E>E_{\rm transition}^{\gamma,\nu}\,.
    \end{array}\right.
\end{equation}
That means for a typical parameter combination of $l_c=10^{11}$~m and $B=0.42$~G, the transition energy for protons is $E_{\rm transition}^{\rm CR}=10^{15}$~eV. This translates (see e.g.\ \cite{a7:beckertjus_multimessenger}  into a transition energy for photons of $E_{\rm transition}^{\gamma}\approx 1/10\cdot E_{\rm transition}^{\rm CR}\approx 10^{14}$~eV. These 100~TeV are currently barely accessible for gamma-ray telescopes, and so the propagation in a purely diffusive regime is reasonable. Purely ballistic transport, however, leads to a spectrum that is too flat, as the steepening by the diffusive escape timescale due to the diffusion tensor is neglected.

For neutrinos, the transition energy is $E_{\rm transition}^{\nu}\approx 1/20\cdot E_{\rm transition}^{\rm CR}\approx 5\cdot 10^{13}$~eV. As the  neutrinos detected by IceCube are in the energy range 10~TeV to a few PeV, this transition region can fall right into the relevant parameter range, so that a combination of ballistic and diffusive propagation needs to be considered. A break in the observed neutrino energy spectrum from a steeper to a flatter behavior is therefore expected in such a scenario. If such a break is observed, it can be used to estimate the parameter combination $B\cdot l_c$. 
To summarize the above result in the context of the propagation of particles in the plasmoids of blazars, we show  that it is of high importance to evaluate the transport regime and adjust it accordingly to the problem under consideration in order to receive reliable results.

%\newpage

%%%%%%%%%%%%%%%%%%%%%%%%%%%%%%%%%%%%%%
\section{The time-domain \label{time:sec}}
%%%%%%%%%%%%%%%%%%%%%%%%%%%%%%%%%%%%%%
The result from the previous section applies to steady-state sources with $\delta n/\delta t\approx 0$, where it is implicitly assumed that all particles have already had the time to reach a steady-state limit in their propagation. However, blazars are highly variable objects and individual flares are often modeled by injecting a high-energy particle population on short timescales.
Particle acceleration time-scales in reconnection events responsible for the blob creation in relativistic sources are  on order of $\tau_\mathrm{acc} \sim E/(qBc^2)$, which is typically much shorter than the escape timescales, see e.g.\ \citet{delvalle2016}, motivating a two-zone scenario where acceleration is typically performed in a first step, propagation afterwards. 

While there is scientific consensus that the description of the transport process of particles in turbulent fields is ensured by the general concept discussed in Section \ref{space:sec} in the limit of infinitely large times, the question arises under which conditions and on which timescale such a limit consideration is appropriate. In this section, criteria are derived for which, in a given plasmoid setup, the diffusive approximation still holds.

The general problem of assuming  diffusive propagation during the initial propagation process, for which the diffusive limit is not yet granted, is expressed in the following points:
\begin{itemize}
\item The solution of the diffusion equation results in a Gaussian spatial distribution of the particles in the plasmoid. However, especially in the initial transport phase, this has the consequence that the particles are granted a non-vanishing probability of reaching positions in the plasmoid, which they cannot reach at the initial time due to their finite speed. 
\item Numerical simulations show a linear increase of the running diffusion coefficient, caused by ballistic particle trajectories until a constant value is reached that is known as the final diffusion coefficient $\kappa$ and used within the numerical and theoretical computations of diffusive transport. The discrepancy between the final diffusion coefficient and the actual running diffusion coefficient approaches zero as the propagation length increases but is significant at the beginning.
\end{itemize}
\subsection{Timescale for transition to diffusive propagation}
Whereas the consideration of particle transport via the diffusion equation and its solution of a Gaussian particle distribution cannot distinguish between the initial, ballistic propagation and the subsequent diffusive propagation, the telegraph equation has recently been attributed this ability \cite{litvinenko2013, litvinenko2013b, litvinenko2015, tautz2016}
\begin{align}
    \frac{\partial f}{\partial t} + \tau \frac{\partial^2f}{\partial t^2} = \kappa \left(\frac{\partial^2f}{\partial x^2} + \frac{\partial^2f}{\partial y^2} + \frac{\partial^2f}{\partial z^2}\right).
\end{align}
The telegraph timescale $\tau$ describes the transition between these two propagation phases. This timescale enables us to make a statement about when the diffusive phase is established and when the description of the particle transport via the diffusion equation is sufficiently accurate. If the initial ballistic phase is neglected, $\tau$ disappears and the telegraph equation turns into the well-known diffusion equation. By neglecting adiabatic focusing, the telegraph timescale yields \cite{litvinenko2013}
\begin{align}
    \tau = \frac{3v}{8\lambda} \int \limits_{-1}^{1}\mathrm{d} \mu \left(\int \limits_0^{\mu} \mathrm{d} \mu^\prime \frac{1-\mu^{\prime 2}}{D_{\mu\mu}(\mu^\prime)}\right)^2\,,
\end{align}
with $D_{\mu\mu}$ being the pitch-angle Fokker-Planck coefficient that, in a negligible background field ($\delta B \gg B_0$), yields \cite{shalchi2009}
\begin{align}
    D_{\mu\mu} = (1-\mu^2) D.
\end{align}
Here, $D$ is the pitch-angle Fokker-Planck coefficient at 90$^{\circ}$. The mean-free path $\lambda_\parallel$ is defined as
\begin{align}
    \lambda_\parallel = \frac{3v}{8} \int \limits_{-1}^{1}\mathrm{d} \mu \frac{(1-\mu^2)^2}{D_{\mu\mu}(\mu)}\,.
\end{align}

\subsection{Quantifying the time needed to achieve certain levels of diffusivity}
Since the diffusion equation assumes diffusive transport at all times and, furthermore, the solution uses the final diffusion coefficient over all timescales, the normalization of the solution remains constant in time:
\begin{align}
    4\pi \int \limits_0^\infty \mathrm{d}r\, r^2 f_\mathrm{diff}(r) = 4\pi \int \limits_0^\infty \mathrm{d} r\,\frac{r^2}{(4\pi t \kappa)^{3/2}} \mathrm{exp}{\left(-\frac{r^2}{4\kappa t}\right)}= 1\,.
\end{align}
The normalization may be interpreted as the fraction of particles participating in diffusion \cite{tautz2016}.

On the other hand, the solution of the isotropic telegraph equation yields
\begin{align}
    f_\mathrm{telegraph}(r,t) &= \frac{e^{-t/2\tau}}{4\pi\kappa^{3/2}} \left[ \frac{\delta \left(t-r\sqrt{\tau/\kappa}\right)}{r/\sqrt{\kappa}} I_0 \left( \frac{1}{2}\sqrt{\frac{t^2}{\tau^2}-\frac{r^2}{\kappa \tau}} \right) \right. \notag \\
    &\quad \left. + \frac{\Theta \left(t/\sqrt{\tau}-r/\sqrt{\kappa}\right)}{2\tau^{3/2}\sqrt{\frac{t^2}{\tau^2}-\frac{r^2}{\kappa \tau}}} I_1\left( \frac{1}{2}\sqrt{\frac{t^2}{\tau^2}-\frac{r^2}{\kappa \tau}} \right) \right].
\end{align}
Here, $\Theta(...)$ is the Heaviside step function and $I_\nu(...)$ is the modified Bessel function.
The norm can be computed individually on each of the two terms of the function:
\begin{align}
    \int f_\mathrm{telegraph}(r,t) \, \mathrm{d} r &= \int \frac{e^{-t/2\tau}}{4\pi\kappa^{3/2}} \left[ \frac{\delta \left(t-r\sqrt{\tau/\kappa}\right)}{r/\sqrt{\kappa}} I_0 \left( \frac{1}{2}\sqrt{\frac{t^2}{\tau^2}-\frac{r^2}{\kappa \tau}} \right) \right] \mathrm{d} r \notag \\
    &\quad + \int \frac{e^{-t/2\tau}}{4\pi\kappa^{3/2}} \left[\frac{\Theta \left(t/\sqrt{\tau}-r/\sqrt{\kappa}\right)}{2\tau^{3/2}\sqrt{\frac{t^2}{\tau^2}-\frac{r^2}{\kappa \tau}}} I_1\left( \frac{1}{2}\sqrt{\frac{t^2}{\tau^2}-\frac{r^2}{\kappa \tau}} \right) \right] \mathrm{d} r\,.
\end{align}
The delta distribution simplifies the first part to $t/\tau$ and the second part can be solved by employing Bessel function integration rules. After tedious calculation, which we omit here, we receive
\begin{align}
    N = 1-\exp\left(-\frac{t_\mathrm{diff,N}}{\tau}\right)\,,
\end{align}
and thus a time-dependent result. In this scenario,  no particles are diffusive at the beginning. With time, the number of diffusive particles increases exponentially until the value for large propagation times approaches the maximum value with all particles in a diffusive state. 

\begin{figure}
\centering
\includegraphics[width=0.8\textwidth]{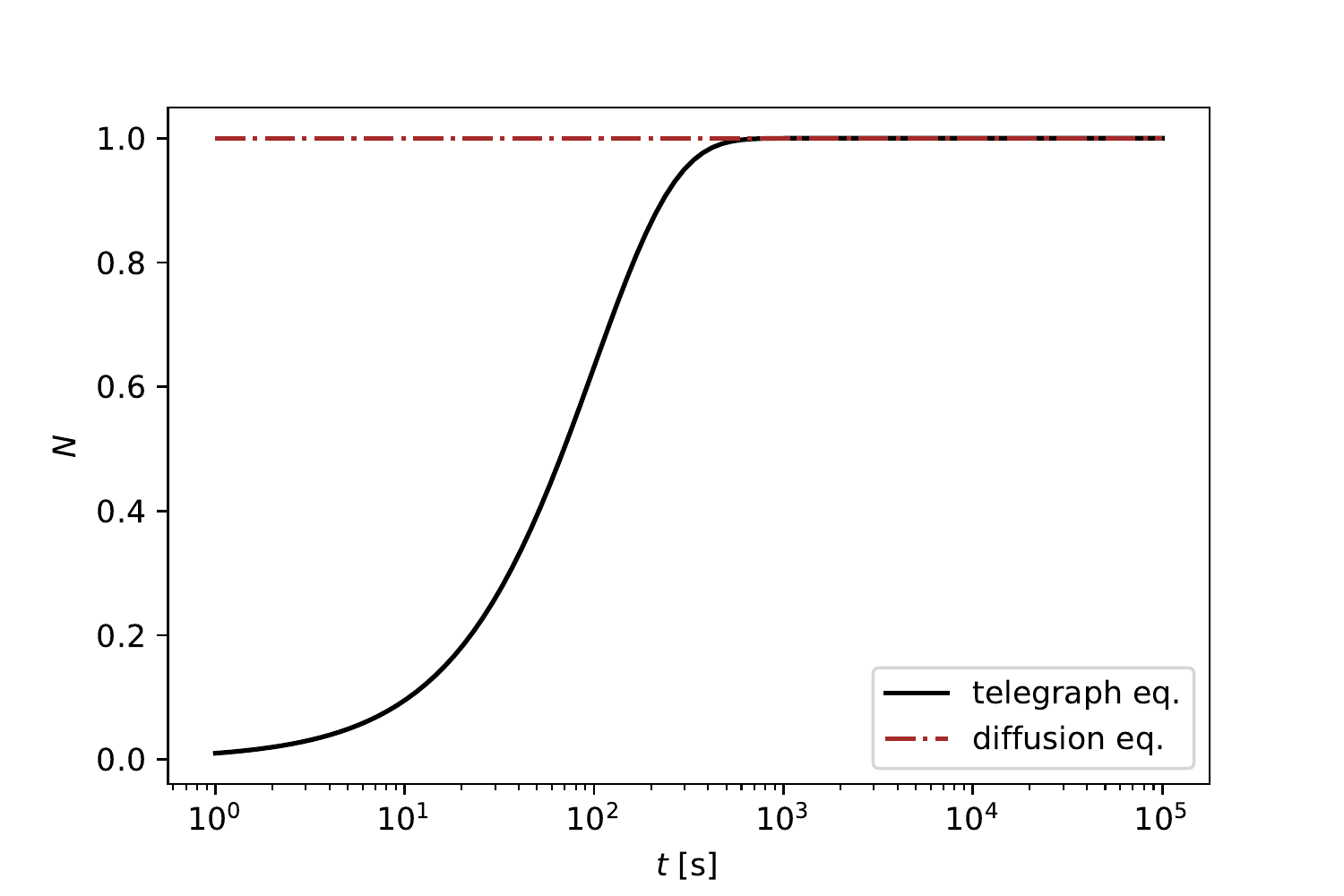}\caption{Comparison of the time evolution of the ratio of particles that are already diffusively propagating on average. The solution of the diffusion equation leads to the fact that particles always propagate diffusively. The solution of the Telegraph equation shows an increase in the diffusively propagating particles with time and an approach to the maximum value.}\label{fig:diffusivity}
\end{figure}

Rearranging the equation leads to a calculation rule for the propagation time required to establish a certain diffusion level:
\begin{align}
    t_\mathrm{diff,N} = -\ln{(1-N)}\tau\,.
    \label{eq:tdiff}
\end{align}
This relation is shown in Fig.\ \ref{fig:diffusivity} in comparison with the constant number of diffusing particles in the case of modeling the transport with the diffusion equation.
The initial phase of a flare of cosmic rays is therefore in a non-diffusive state until the steady-state limit is reached as shown in Fig.\ \ref{fig:diffusivity}. Note that the derivations made here for a purely turbulent field also apply to the generalized case of an additional directional magnetic field component, since the timescales for reaching the diffusive phase during transport parallel and perpendicular to the directional background field are identical for a large parameter space \cite{Reichherzer2021BP}. 

In the following subsection, we will evaluate this scenario for the conditions in a plasmoid.

\subsection{Conditions for plasmoid settings}

In the following, the critical time to reach diffusive propagation is expressed as a function of the blob properties and the particle energy. The resulting estimates give an overview of the expected type of propagation for special parameter combinations.
For this purpose, we start with the definition of the timescale connected to the mean free path $\lambda$ of the particle,
\begin{align}
    \tau = \frac{\lambda}{v} = \frac{3\kappa}{v^2}\,,
\end{align}
where $\lambda=3\kappa/v$ is expressed as a function of the diffusion coefficient.
%exploiting in the last transformation step the relationship between the diffusion coefficient and the mean-free path length. 
We consider the case of isotropic turbulence without background field here, in which case  Bohm diffusion applies with a linear energy dependence in the resonant-scattering regime, $\kappa \propto E$. In the quasi-ballistic regime, the diffusion coefficient becomes   $\kappa\propto E^2$. The transition between the two regimes is given at a reduced rigidity of $\rho\approx 5/(2\pi)$, so that the diffusion coefficient can be written as
\begin{equation}
\kappa = \kappa_0 \left(\frac{\rho}{\rho_0} \right)^\delta \mathrm{ with }\begin{cases}
\delta = 1&    $for $\rho \lesssim 5/(2\pi) \\
\delta = 2&    $for $\rho \gg 1
\end{cases}\,.
\end{equation}

\begin{figure}[h]
\centering
\includegraphics[width=0.8\textwidth]{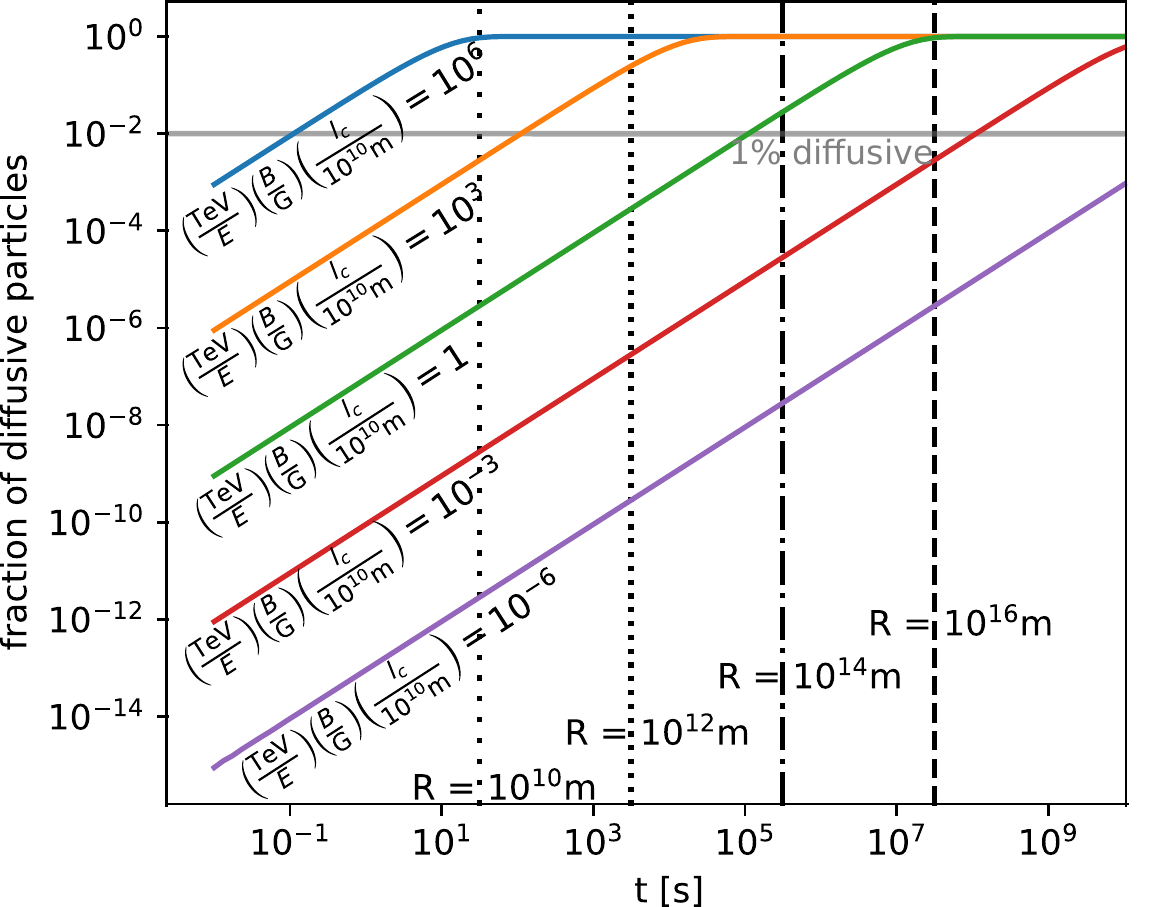}\caption{Fraction of particles that are diffusive as a function of propagation time for different blob parameters and particle energies. The vertical lines illustrate the time required for ballistic particle propagation to traverse the respective blob radii. }\label{fig:diffusive_particles}
\end{figure}

It follows for the timescale
\begin{equation}\label{eq:tau_vs_rho}
\tau = \frac{3\kappa_0}{v^2}  \left(\frac{\rho}{\rho_0} \right)^\delta \mathrm{ with }\begin{cases}
\delta = 1&    $for $\rho \lesssim 1 \\
\delta = 2&    $for $\rho \gg 1
\end{cases}\,,
\end{equation}
finally leading to an expression for the 
propagation time required to reach a certain diffusion level $N$ (from Eq.\ (\ref{eq:tdiff}))
\begin{equation}
t_{\mathrm{diff},N}= -\ln{(1-N)} \frac{3\kappa_0}{v^2}  \left(\frac{2\pi \rho}{5} \right)^\delta \mathrm{ with }\begin{cases}
\delta = 1&    $for $\rho \lesssim 1 \\
\delta = 2&    $for $\rho \gg 1
\end{cases}\,.
\label{eq:rho}
\end{equation}
By inserting the definition of the reduced rigidity, this relation can be expressed as functions of $E$, $B$ and $l_\mathrm{c}$:
\begin{equation}
t_{\mathrm{diff},N}= -\ln{(1-N)} \frac{3\kappa_0}{v^2}  \left(\frac{2\pi E}{5qcBl_\mathrm{c}} \right)^\delta \mathrm{ with }\begin{cases}
\delta = 1&    $for $\rho \lesssim 1 \\
\delta = 2&    $for $\rho \gg 1
\end{cases}\,.
\end{equation}
The time $t_{\mathrm{diff},N}$ needed for the fraction $N$ of the particles to be diffusive depends on the parameters $E$, $B$ and $l_\mathrm{c}$. 
Figure \ref{fig:diffusive_particles} shows this condition for different plasmoid parameters. The figure shows the influence of the particle energy, the magnetic field properties, and the trajectory on the fraction of already diffusively propagating particles. The vertical lines indicate the timescales required for particles on ballistic trajectories to travel one plasmoid radius. This timescale has the same order as typical escape times of charged particles during initial ballistic propagation. If there is not yet a significant fraction of diffusive particles at the vertical lines for the respective plasmoid radii, the particles must be considered completely ballistic. For example, charged particles with $E=1\,$TeV, $B=1\,$G, and $l_\mathrm{c}=10^{10}\,$m can be treated diffusively in plasmoids with $R=10^{16}\,$m, but must be treated via equation-of-motion approaches at smaller radii such as $R=10^{14}\,$m, $R=10^{12}\,$m, and $R=10^{10}\,$m.

%%%%%%%%%%%%%%%%%%%%%%%%%%%%
\section{Simulations: ballistic VS.\ diffusive simulation results \label{sec:simulation}}
\begin{figure}[h]
\centering
\includegraphics[width=0.8\textwidth]{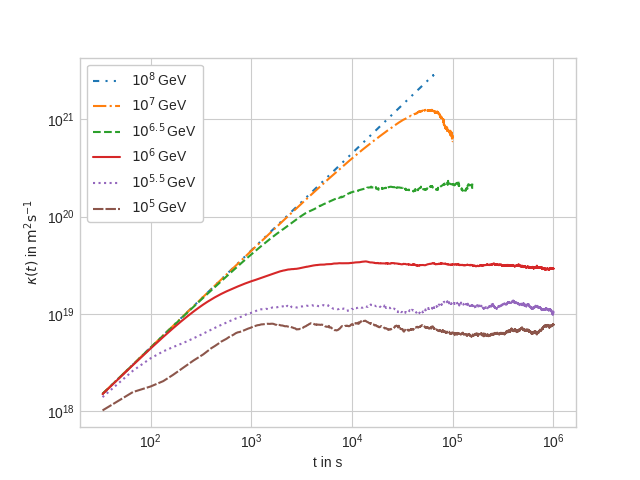}\caption{Running diffusion coefficient for energies from $10^{5}$~GeV to $10^{8}$~GeV. A plateau is built up for energies between $10^{5}$~GeV~$<E<10^{6}$~GeV. Toward higher energies, the coefficient breaks off as the particles leave the plasmoid before they can reach the steady-state diffusion coefficient. This is consistent with the calculated energy for a transition between a ballistic and diffusive behavior at $10^{6}$~GeV.
%\Julia{@Marcel: bitte Linien \"uber Strichelungen unterscheiden, damit sie auch s/w lesbar sind} 
}\label{fig:runningdiffusion}
\end{figure}

\begin{figure}[h]
\centering
\includegraphics[width=0.8\textwidth]{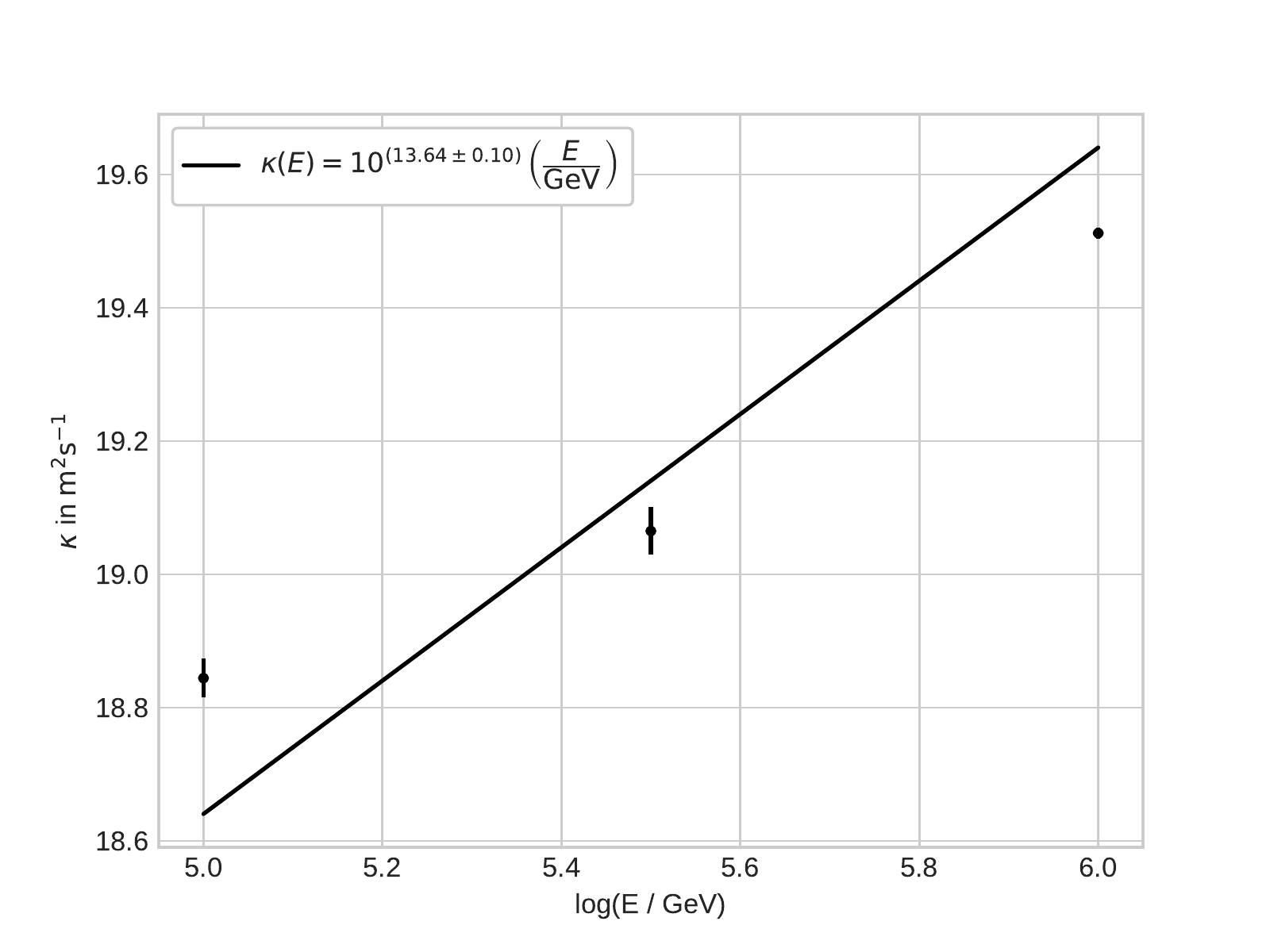}\caption{Steady-state diffusion coefficient as a function of energy for particles between $10^{5}$~GeV and $10^{6}$~GeV. A linear regression for the function $\kappa(E)=\kappa_0\cdot (E/$GeV$)$ is performed. The linear behavior with energy is based on the assumption that Bohm diffusion is dominant in the purely turbulent field.  }\label{fig:kappa}
\end{figure}

%%%%%%%%%%%%%%%%%%%%%%%%%%
In this section, we investigate the effect of ballistic vs.\ diffusive propagation by performing simulations of cosmic-ray transport in the plasmoid of an AGN traveling along the jet axis. We use the code developed in \cite{ownpub_a7:model2020}. Here, we switch off interactions with photon and gas targets in order to focus on the effects coming from cosmic-ray propagation, but otherwise follow the procedure described in \cite{ownpub_a7:model2020}. The parameter set used in the simulation is summarized in Table \ref{tab:globalparameters}. %\Julia{@Marcel: Tabelle der Einstellungen, vergleich Marios Papier, zB Plasmoid-Gr\"o\ss e, Korrelationsl\"ange, ...} %\Marcel{Soll ich dann noch eine weitere Tabelle erstellen, in denen ich die einzelnen Simulationen nach Parametern auflöse? An sich habe ich an den Basisparametern für diese Simulationen im Vergleich zu Mario nicht viel geändert.}. 
In addition, we not only perform simulations with the equation-of-motion, but apply diffusive propagation by using the module DiffusionSDE in CRPropa 3.1 \cite{merten2017}, which solves the transport equation with a diffusion term $\kappa \Delta n$. In order to have a quantitative comparison, we first need to calculate the diffusion coefficient $\kappa$ as an input to the diffusive simulation from the ballistic part. We do this for energies from $10^{5}$~GeV up to $10^{8}$~GeV. Here, we apply the Taylor Green Kubo (TGK) formalism as described in \cite{reichherzer2020} (see also references therein). We choose the simulation parameters in Table \ref{tab:globalparameters} to minimize numerical errors such as the interpolation effect of turbulence \cite{Schlegel2020, Reichherzer:2021a}. Figure \ref{fig:runningdiffusion} shows the result of the running diffusion coefficient $\kappa(t)$. For low energies, i.e.\ $10^{5}$~GeV (brown), $10^{5.5}$~GeV (purple), $10^{6}$~GeV (red), and $10^{6.5}$~GeV (green), particles reach a plateau after an initial ballistic phase as described in Section \ref{time:sec}, which can be used as the steady-state diffusion coefficient. For larger energies ($10^{7}$~GeV, orange, and $10^{8}$~GeV, blue), such a convergence is not observed. The reason is that the particles leave the plasmoid before being able to reach a steady-state diffusion limit. We therefore use the first three values of the steady-state diffusion coefficient to determine the energy dependence $\kappa(E)$. In Fig.\ \ref{fig:kappa}, the values averaged from the plateau in Fig.\ \ref{fig:runningdiffusion} are shown with the corresponding error bars. In the simulation setup, the magnetic field is a purely turbulent one. That means Bohm diffusion is at work, and the energy dependence is expected to be $\kappa=\kappa_0\cdot (E/$~GeV$)$, see e.g.\ \citep{a7:beckertjus_merten2020}. We therefore perform a linear regression and find $\kappa_0=10^{13.64\pm0.10}$. For our simulations, we use the calculated values for the three energies where this was possible in a reliable way ($10^{5}$~GeV to $10^{6}$~GeV). For larger values, we use the result from the linear regression to determine the diffusion coefficient. From our findings in Sections \ref{space:sec}, we expect the ballistic and diffusive flares to provide approximately the same results until the transition energy is reached according to Eq.\ (\ref{eq:for_Hillas_plot}), for our set of parameters (see Table \ref{tab:globalparameters}), this is at around $10^{6}$~GeV. From our results in Section \ref{time:sec} we expect a deviation between the diffusive and ballistic approach that is largest at small times and the two approaches should converge toward large times, when the steady-state diffusion coefficient is reached. As can be seen from Fig.\ \ref{fig:runningdiffusion}, this effect is energy dependent and for low energies ($10^{5}$~GeV), the diffusive steady-state is reached at around $10^{3}$~s, while it takes $\gg 10^{4}$~s for the highest energies ($E>10^{7}$~GeV).

	\begin{table}
		\begin{tabular}{lc}
\hline
			Parameter                        &      Value       \\ \hline
			Proton energy $E_\mathrm{p}$                   & $10^5\,\mathrm{GeV}\,-\,10^8\,\mathrm{GeV}$  \\
			Plasmoid radius  $R$                  & $10^{13}\,\mathrm{m}$ \\
			Plasmoid Lorentz factor $\Gamma$         &       $10$       \\
			Magnetic field: Initial RMS value $B_0$ &    $1\,\mathrm{G}$    \\
			Magnetic field: Turbulence \& spectral index $\alpha$  & Kolmogorov-type, $\alpha=-5/3$   \\
			Magnetic field: Correlation length  $l_c$  & $10^{11}\,$m \\
			Magnetic field: Grid points & $\left(512\right)^3$\\
			Magnetic field: Spacing & $R/256$ \\
			Propagation module (CRPropa intern): Ballistic & PropagationBP \\
			Propagation module (CRPropa intern): Diffusive & DiffusionSDE \\
			Propagation step size & $10^{-3}R$\\\hline
		\end{tabular}
		\caption{Simulation parameters, given in the rest frame of the plasmoid. All simulations are performed in a modified of CRPropa 3.1 as presented in Hörbe et al.\ in \cite{ownpub_a7:model2020} with further additions made for this paper as described above. 
		}
		\label{tab:globalparameters}
	\end{table}

\begin{figure}[h]
\centering
\includegraphics[width=0.8\textwidth]{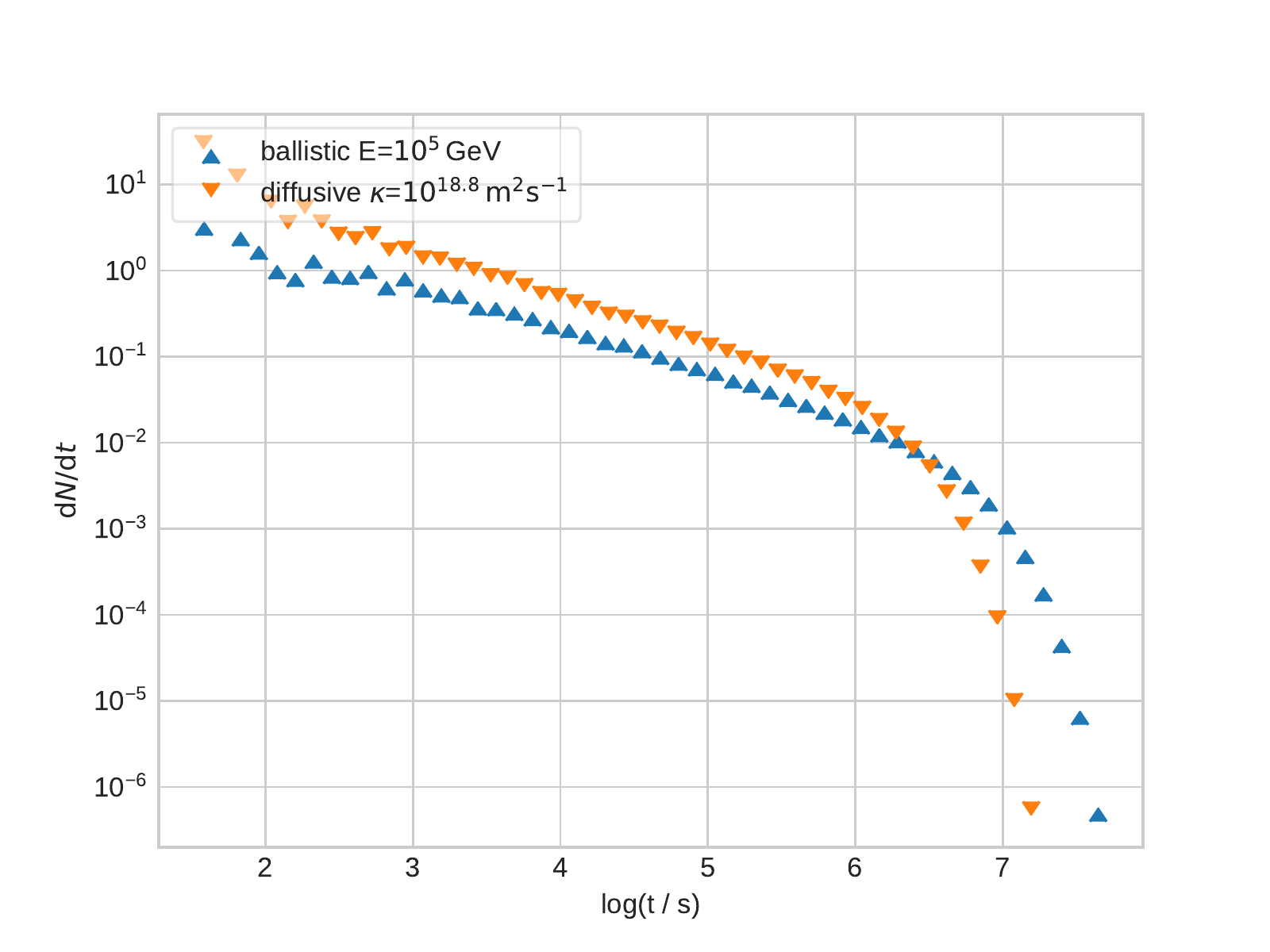}\caption{Cosmic-ray flare ($E=10^{5}$~GeV) from a blazar in the diffusive propagation model (orange downward triangle) and in comparison in the ballistic propagation model (blue upward triangle) as differential particle number per unit time $\mathrm{d}N/\mathrm{d}t$ over time. The total number of particles injected into the simulation is $N_{\rm inj}=10^5$.
}\label{fig:flare_low_energy}
\end{figure}

Figure \ref{fig:flare_low_energy} shows the flaring behavior for a monochromatic energy flare at $E=10^{5}$~GeV. 
The diffusive description (orange downward triangle) shows an especially large enhancement with respect to the equation-of-motion approach (blue upward triangle) at early times below $10^{3}$~s, in accordance with our findings in Section \ref{time:sec}. The behavior of $\mathrm{d}N/\mathrm{d}t$ is similar for both approaches, with a small shift that can be explained by the uncertainties in our numerical determination of the diffusion coefficient used for the diffusive approach.
In the diffusive transport regime for a constant diffusion coefficient, we expect $\langle \Delta x \rangle \propto t^{1/2}$. This results in the differential particle number $\mathrm{d}N/\mathrm{d}t \propto t^{-1/2}$ of escaping particles. 
Note that, due to the steady escape, the decrease in the number of remaining particles in the plasmoid leads to a strong cut-off at large times. Since, in the diffusive approach, more particles initially leave the plasmoid, the particle density in the plasmoid is lower than in the equation-of-motion approach, so that an earlier cut-off is visible. 

\begin{figure}[h]
\centering
\includegraphics[width=0.8\textwidth]{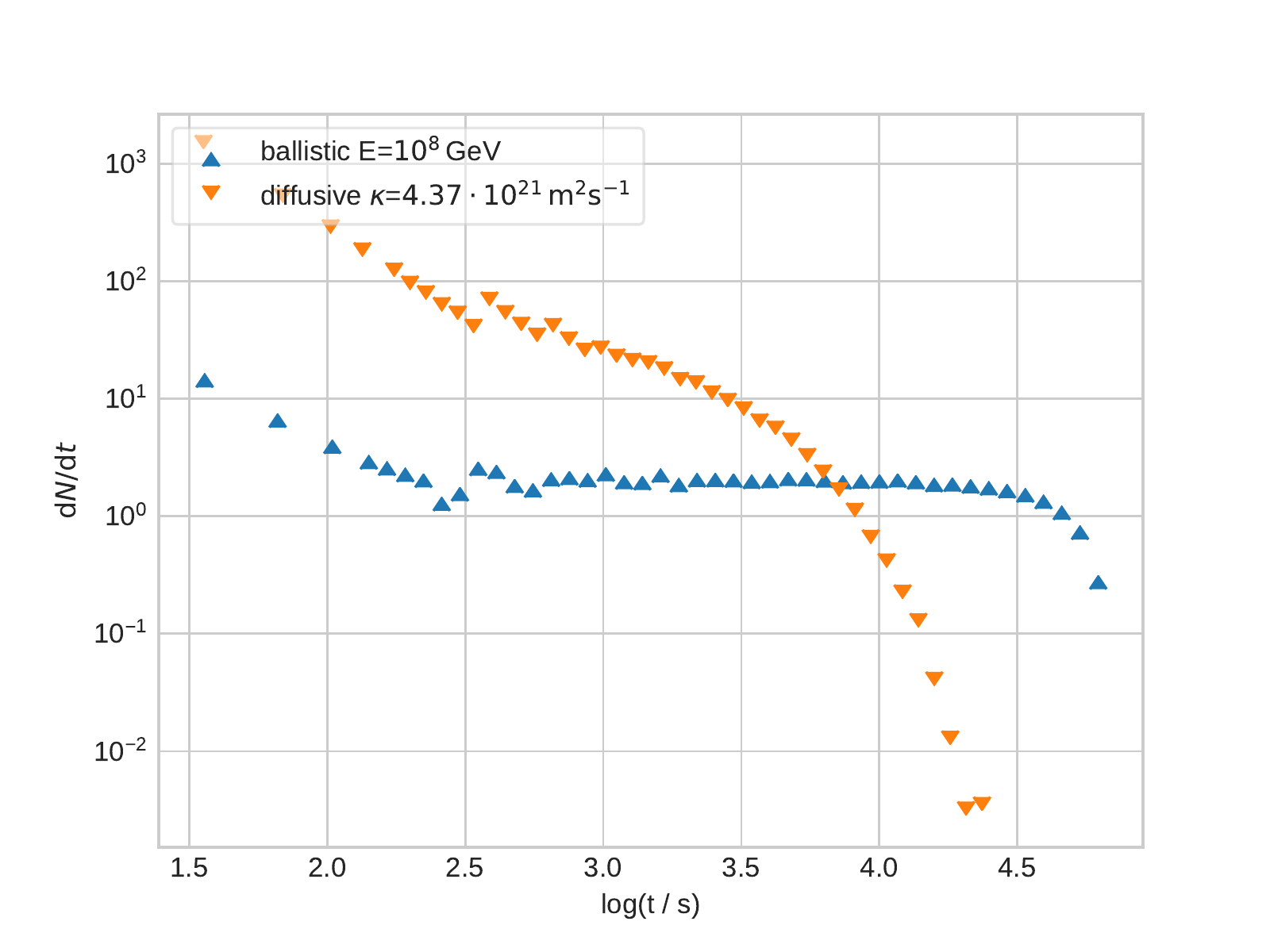}\caption{Cosmic-ray flare ($E=10^{8}$~GeV) from a blazar in the diffusive propagation model (orange downward triangle) and in comparison in the ballistic propagation model (blue upward triangle) as differential particle number per unit time $\mathrm{d}N/\mathrm{d}t$ over time. The number of injected particles is again $N_{\rm inj}=10^5$.  }\label{fig:flare_high_energy}
\end{figure}

Figure \ref{fig:flare_high_energy} shows the flare for diffusive (again orange downward triangle) and equation-of-motion (again blue upward triangle) behavior at $10^{8}$~GeV.
Here, there is a very clear difference between the two flares, with the diffusive approach yielding a dominant contribution at early times. In contrast to the diffusive regime with $\mathrm{d}N/\mathrm{d}t \propto t^{-1/2}$, we expect a constant differential particle number for the ballistic transport regime with $\langle \Delta x\rangle \propto t$. The initial slight drop for the equation-of-motion may be explained by statistical deviations from the initial homogeneous particle distribution in the plasmoid, especially when slightly more particles are in the outer spheres of the plasmoid at $t=0$. Since, in the diffusive case, significantly more particles initially leave the plasmoid, the particle density in the plasmoid is much lower than in the equation-of-motion approach, so that a cut-off is visible much earlier in the diffusive approach. Thus, these test simulations emphasize the importance of propagating the particles in the proper transport regime. Only a thorough analysis of the transport properties will lead to a prediction that can be compared to the observation of non-thermal emission from blazars.

\section{Conclusions \label{conclusions:sec}}
In this paper, we investigate the propagation regimes in plasmoids of blazars as sources of high-energy cosmic rays, which in turn become emitters of high-energy gamma-rays and neutrinos. To explain the spectral energy distributions and lightcurves of this high-energy emission, we show that it is necessary to distinguish between the different energy and time regimes of ballistic and diffusive transport. At early times and at high energies, the particles are still in the ballistic regime. At late times or in scenarios for which the injection of high-energy cosmic rays is significantly longer than $\tau\gg R/c$, the diffusive approach needs to be applied. The details of this transport modeling have an impact on both the spectral energy distribution, and on the temporal evolution of a flare. For the energy behavior, the diffusive part of the spectrum is steepened by the diffusion timescale which is dominated by the diffusion coefficient. The ballistic part, on the other hand, is connected to an energy-independent escape timescale, thus leading to an emission spectrum close to the acceleration spectrum. When looking at the flaring behavior, it has been shown that diffusive approach and transport with the equation-of-motion approach yield about the same result at low energies around $10^{5}$~GeV, where the diffusive approach is accurate at times above $\sim 10^{3}$~s. That means that if the equation-of-motion approach is performed with the correct parameter setting, the same result is expected at times larger than $10^{3}$~s, which we can confirm within a factor of $\sim 2$. When approximating this behavior with an escape timescale, the diffusive timescale needs to be applied.

At large energies ($10^{8}$~GeV), we can show that the diffusive and equation-of-motion approaches lead to very different flaring behaviors. Here, only the equation-of-motion approach yields a correct result, as it can reproduce the ballistic behavior. It can be approximated in a transport equation approach by applying an energy energy-independent (ballistic) escape time.

%%%%%%%%%%%%%%%%%%%%%%%%%%%%%%%%%%%%%%%%%%
\authorcontributions{Conceptualization, J.B.T.; methodology, all; writing---original draft preparation, J.B.T., P.R., I.J.; writing---review and editing, all; visualization, P.R., I.J., M.S.; supervision, J.B.T., F.S.; funding acquisition, J.B.T., W.R., F.S. All authors have read and agreed to the published version of the manuscript.}

\funding{This research was funded by the German Science Foundation DFG via the Collaborative Research Center SFB1491 "Cosmic Interacting Matters - From Source to Signal". Further funding was received from the DFG via the grant \textit{Multi-messenger probe of Cosmic Ray Origins (MICRO)}, grant numbers TJ\,62/8-1. We would also like to thank the Research Department for Plasmas with Complex Interactions for support.}

\dataavailability{Data presented in this article can be made available upon request.} 

\acknowledgments{
J.B.T.\ and W.R.\ would like to use this opportunity to thank Reinhard Schlickeiser for the long, pleasant journey through physics, administration, and the Ruhr area during the past decades. We hope the journey will continue for long - even if its path might shift in its character.

We would like to thank Rainer Grauer for discussions on the plasma physics of blazars, particularly concerning the launching and evolution of plasmoids and their plasma parameters. We also thank Imre Bartos, Peter Biermann, Anna Franckowiak, Francis Halzen, Emma Kun, and Walter Winter for discussions on the modeling of non-thermal blazar emission}

\conflictsofinterest{The authors declare no conflict of interest.} 

%%%%%%%%%%%%%%%%%%%%%%%%%%%%%%%%%%%%%%%%%%
\begin{adjustwidth}{-\extralength}{0cm}

\reftitle{References}

\bibliography{literature}

%=====================================
% References, variant B: internal bibliography
%=====================================

\end{adjustwidth}
\end{document}